\documentclass[pre,superscriptaddress,twocolumn,nofootinbib,floatfix]{revtex4}
\usepackage{amsmath,amsfonts,amssymb}
\usepackage{graphicx}
\usepackage{hyperref}
\usepackage{verbatim}

\newcommand{\bi}{\begin{itemize}} \newcommand{\ei}{\end{itemize}}
 
\newcommand{\bn}{\begin{enumerate}} \newcommand{\en}{\end{enumerate}}
\newcommand{\ba}{\begin{array}} \newcommand{\ea}{\end{array}}
\newcommand{\bc}{}
\newcommand{\be}{\begin{equation}} \newcommand{\ee}{\end{equation}}
\newcommand{\bex}{\begin{equation*}} \newcommand{\eex}{\end{equation*}}
\newcommand{\bea}{\begin{eqnarray}} \newcommand{\eea}{\end{eqnarray}}
\newcommand{\beax}{\begin{eqnarray*}} \newcommand{\eeax}{\end{eqnarray*}}

\newcommand{\btp}[6]{\begin{tikzpicture}[smooth, domain=#2:#3]  \draw[->](#2  ,0) -- (#3,0 ) node[right]{#1}; \draw[->](0,#5) -- (0,#6) node[above] {#4}; \clip (#2,#5) rectangle(#3,#6);}

\begin{document}

\renewcommand*{\thefootnote}{\fnsymbol{footnote}}

\title{Focal conic flower textures at curved interfaces}
\author{Daniel A. Beller}
\affiliation{Department of Physics and Astronomy, University of Pennsylvania, Philadelphia, Pennsylvania 19104, USA}
\author{Mohamed A. Gharbi}
\affiliation{Department of Physics and Astronomy, University of Pennsylvania, Philadelphia, Pennsylvania 19104, USA}
\affiliation{Department of Chemical and Biomolecular Engineering, University of Pennsylvania, Philadelphia, Pennsylvania 19104, USA}
\affiliation{Department of Materials Science and Engineering, University of Pennsylvania, Philadelphia, PA 19104, USA}
\author{Apiradee Honglawan}
\affiliation{Department of Chemical and Biomolecular Engineering, University of Pennsylvania, Philadelphia, Pennsylvania 19104, USA}
\author{Kathleen J. Stebe}
\affiliation{Department of Chemical and Biomolecular Engineering, University of Pennsylvania, Philadelphia, Pennsylvania 19104, USA}
\author{Shu Yang}
\affiliation{Department of Chemical and Biomolecular Engineering, University of Pennsylvania, Philadelphia, Pennsylvania 19104, USA}
\affiliation{Department of Materials Science and Engineering, University of Pennsylvania, Philadelphia, PA 19104, USA}
\author{Randall D. Kamien\thanks{Electronic address: kamien@physics.upenn.edu}}
\affiliation{Department of Physics and Astronomy, University of Pennsylvania, Philadelphia, Pennsylvania 19104, USA}


\date{\today
}

\begin{abstract}Focal conic domains (FCDs) in smectic-A liquid crystals have drawn much attention both for their exquisitely structured internal form and for their ability to direct the assembly of micro- and nanomaterials in a variety of patterns. A key to directing FCD assembly is control over  the eccentricity of the domain. Here, we demonstrate a new paradigm for creating spatially varying FCD eccentricity by confining a hybrid-aligned smectic with curved interfaces. In particular, we manipulate interface behavior with colloidal particles in order to experimentally produce two examples of what has recently been dubbed the flower texture {[}C. Meyer et al., {Materials}, vol. 2, pp. 499-513, 2009{]}, where the focal hyperbol{\ae}  diverge radially outward from the center of the texture, rather than inward as in the canonical \textit{eventail} or fan texture. We explain how this unconventional assembly can arise from appropriately curved interfaces. Finally, we present a model for this system that applies the law of corresponding cones, showing how FCDs may be embedded smoothly within a ``background texture" of large FCDs and concentric spherical layers, in a manner consistent with the qualitative features of the smectic flower. Such understanding could potentially lead to disruptive liquid crystal technologies beyond displays, including patterning, smart surfaces, microlens arrays, sensors and nanomanufacturing. \end{abstract}

\maketitle

Exploiting the elasticity and surface anchoring of liquid crystals has opened up a new world of self-organizing behaviors in which liquid crystalline defects, rather than individual molecules, are components of self-assembly \cite{Musevic2006, Ravnik2011, nych2013assembly}. In the smectic-A liquid crystal (SmA LC) phase, geometrically robust layer arrangements called focal conic domains (FCDs), organized around a pair of defect curves, have drawn much recent attention for their assembly into ordered arrangements over large areas in hybrid-aligned cells \cite{yoon2007internal, Honglawan2011, zappone2012periodic}. Ordered arrays of FCDs have been investigated for a variety of technological applications \cite{Kim2011}, such as regular arrays of trapped colloids \cite{ yoon2007internal}, superhydrophobic surfaces \cite{kim2009fabrication}, optically selective photomasks \cite{kim2010optically}, microlens arrays \cite{kim2010fabrication}, and soft lithography templates \cite{kim2010self}. A common theme in much research on self-assembly in LCs is the sensitive dependence of the assembly behavior on non-trivial boundary geometry, such as colloid shape \cite{gharbi2013microbullet, senyuk2012topological, lapointe2009shape, dontabhaktuni2012shape} and substrate topography \cite{Honglawan2011, Honglawan2013, choi2004ordered, kim2009confined}. 

Even when not organized in a lattice, FCDs exhibit a high level of geometric organization as seen in the arrangement of their focal curve pairs, which are conjugate conic sections: an ellipse and a hyperbola (or two parabol\ae, a case that we will not study here). Typically, groups of FCDs spontaneously assemble into the so-called fan texture, with the hyperbol\ae~all intersecting at a single point. Friedel \cite{Friedel1922}, in a theory supplemented by later authors \cite{bragg1934liquid, Sethna1982, Lavrentovich1986, kleman2000grain}, explained the fan texture by positing the {\em law of corresponding cones} (LCC), in which the smectic layers smoothly join together neighboring FCDs across conical boundary surfaces. These geometrical rules suggest a route to targeted assembly of FCDs with vastly increased sophistication as a result of nonzero eccentricity of the ellipse in the conjugate pair \cite{Honglawan2013, zappone2012periodic, ohzono2012focal}.  

\begin{figure}
	   \centering
        	\includegraphics[width=0.5 \textwidth]{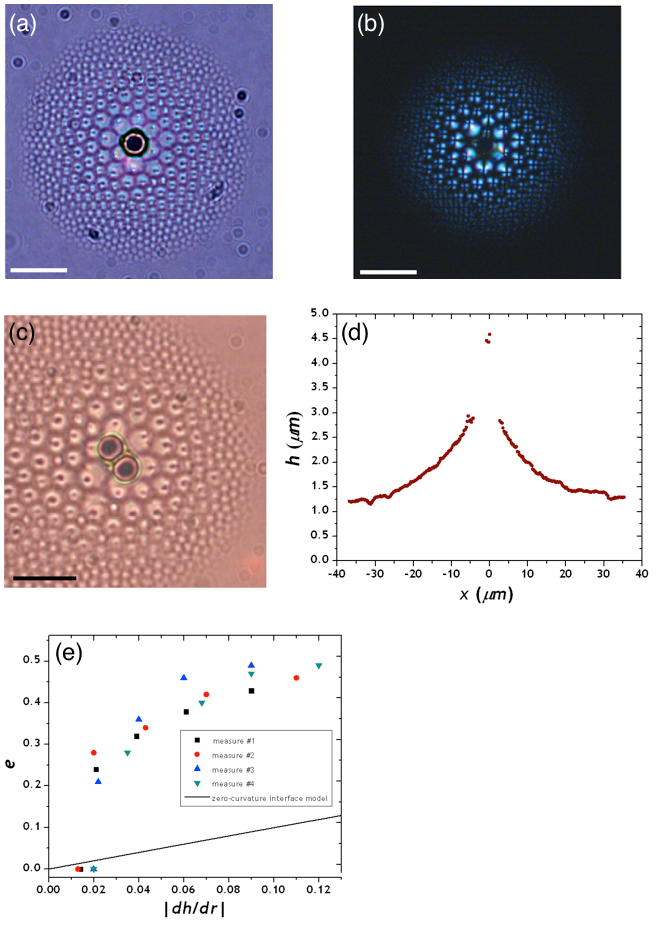}                 
        \caption{(Color) System A.  (a-b) Smectic flower texture organized around a single colloid with homeotropic anchoring in response to the distortion of the LC-air interface produced by pinning at the colloid boundary. (a) Bright field. (b) Polarized optical microscopy. (c)  Flower texture organized around a colloidal dimer. (d) Interferometric measurement of the LC-air interface profile around one colloid: smectic film thickness as a function of distance from the colloid center. (e) FCD eccentricity vs. magnitude of local slope of the smectic-air interface in the radial direction, measured above the middle of ellipses in four different radial directions. Solid line is the eccentricity given in Equation \ref{eq:KLe} corresponding to the limit of zero interfacial curvature. All scale bars are 10 $\mu$m.}\label{MG_flower1}
\end{figure}

A supreme example of FCD self-organization with nonzero eccentricity is the ``flower texture" in a smectic droplet reported in Ref. \cite{meyer2009focal}. There, many FCDs pack with their ellipse long axes oriented radially  from a common point $P$. However, the foci of the ellipses that are pierced by the hyperbol\ae, seen easily in bright field microscopy, are on the ``far side'' of the ellipse -- unlike in the fan texture where the  hyperbol\ae~converge, in the flower they diverge away from $P$ with no obvious intersection, apparently violating Friedel's LCC.

In this article, we show that such packings of FCDs with diverging hyperbol\ae~can be obtained by designing hybrid anchoring conditions such that one boundary is (approximately) a surface of revolution with negative slope in the radial outward direction. More generally, we demonstrate that curved interfaces provide a way to promote spatially varying FCD eccentricity, leading to complex patterns which could guide the assembly of technologically important materials, such as colloids, nanoparticles, and quantum dots, for novel metamaterials, sensors, optoelectronic devices, and solar cells \cite{stebe2009oriented, trindade2001nanocrystalline, lee2002ordering, garcia2006quantum}. Further, thanks to the lensing properties of individual FCDs \cite{kim2010fabrication}, arrays of FCDs organized radially as in the flower texture could efficiently focus light toward a central point, where the {\em virtual} (unphysical) branches of the hyperbol\ae~intersect, for optical and photovoltaic applications.   

\begin{figure}
	\centering
	\includegraphics[width=0.5 \textwidth]{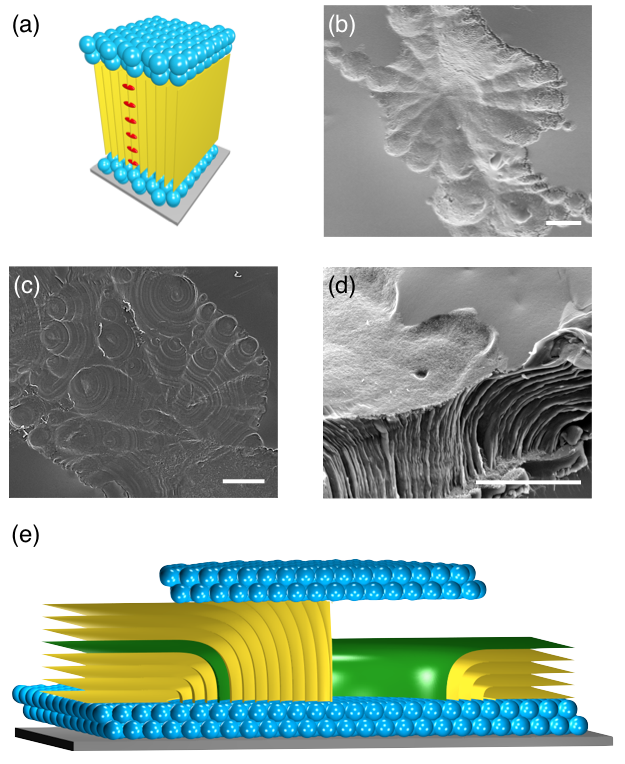}
	\caption{(Color) System B. (a) Schematic illustration of smectic phase (yellow layers, with representative rod-like molecules in red) between two interfaces covered with silica nanoparticles, shown in blue. (b) Smectic flower texture in a nanoparticle-covered region surrounded by a nanoparticle-depleted region.  (c) A less equilibrated smectic flower texture. The smectic layer arrangement at the top interface is visible in the arrangement of the nanoparticles. (d) Cross-section of smectic liquid crystal at the boundary between a nanoparticle-covered region and a nanoparticle-depleted region. (e) Schematic illustration of the geometry of (d), with an arbitrarily chosen bent layer colored green to represent the analog of System A's curved homeotropic interface. All scale bars are 10 $\mu$m.}
	\label{AH_FCDs}
\end{figure}

We present two examples of smectic flower textures obtained from different material systems.  In System A, the smectic-air interface is deformed by pinning at the boundary of a large colloidal inclusion, resulting in a flower texture with FCDs organized radially around the colloid. In System B, a SmA LC is placed on a substrate promoting degenerate planar anchoring, and the air interface, which imposes homeotropic anchoring, is partly replaced by a fluorosilane modified layer of $\mathrm{SiO_2}$ nanoparticles that instead impose degenerate planar anchoring on the LC (Fig.~\ref{AH_FCDs}a). The smectic layers tilt toward the boundary between the nanoparticle-covered and nanoparticle-depleted regions, and FCDs of varying eccentricity interpose between these tilted layers and the nanoparticle interface. In both systems, the key geometric feature is a mismatch in orientation between the interface with degenerate planar anchoring and the smectic layers at the opposite boundary. Finally, we provide a theoretical model applying the LCC to the flower texture. By geometrically constructing a ``background texture''  that approximates an arbitrary homeotropic interface profile, we show that FCDs of nonzero eccentricity can be smoothly embedded such that their hyperbol\ae~extend radially outward, without violating the LCC.

In both System A and System B, smectic flower textures are observed in thin smectic films subjected to (effectively) hybrid anchoring conditions. Figure \ref{MG_flower1} shows an example of such a texture in System A, in which flower textures assemble around a large colloidal inclusion. The average smectic thickness is smaller than the critical thickness $h_c$ below which FCDs cost greater energy than homeotropically-aligned layers  \cite{kim2009confined}. However, pinning of the LC-air interface at the colloid increases the thickness locally above $h_c$, so that FCDs form near the colloid. The film thickness, and thus the typical domain size, decrease with increasing distance from the colloid. Under bright field microscopy (Fig.~\ref{MG_flower1}a), the nonzero eccentricity of the FCDs is apparent both from the elongation of the ellipses and from the off-center dots marking the termination of the hyperbol\ae. As in Ref. \cite{meyer2009focal}, the hyperbol\ae~are oriented radially outward from the center, in contrast to typical fan textures where the hyperbol\ae~converge to a central point. This organization is confirmed by polarized optical microscopy (Fig.~\ref{MG_flower1}b), which reveals dark crosses shifted off of the ellipse centers away from the colloid. The interfacial deformation created by the colloid leads to capillary attraction between nearby colloids to minimize the excess free energy caused by the overlap of deformations in the LC-air interface. Figure (\ref{MG_flower1}c) shows a colloidal dimer with nearby FCDs. 

Nonzero eccentricity is correlated with nonzero slope of the LC-air interface due to surface pinning at the colloid, which satisfies wetting conditions at particle surfaces as described by the Young equation \cite{Stebe2004wetting}. The slope in the radial outward direction decreases from a maximum at the colloid to zero asymptotically, as shown in Fig.~\ref{MG_flower1}d, where the profile of the LC-air interface around the inclusion, measured using scanning white-light interferometry (SWLI), is represented. Accordingly, the FCDs nearest to the colloid have the highest eccentricity, while those far away have nearly zero eccentricity  (Fig.~\ref{MG_flower1}e). Thus, FCD eccentricity is controlled by the orientation mismatch between the horizontal substrate, which imposes degenerate planar anchoring, and the locally tilted LC-air interface, which imposes homeotropic anchoring. 

\begin{figure}	
	\centering
	\includegraphics[width=.5\textwidth]{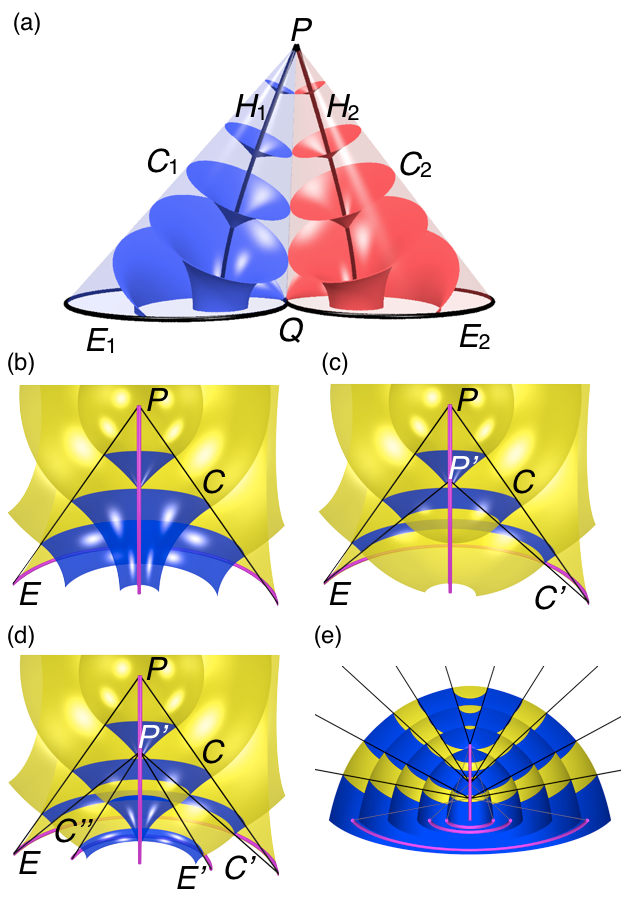}			
	\caption{ (Color) Geometric construction of a background texture. (a) The law of corresponding cones (LCC) in the traditional case of converging hyperbol\ae, with focal curves in black. (b-e) Construction of a complex background texture for the flower texture in accordance with the LCC. TFCDs are shown in blue, while concentric spherical regions are shown in yellow. Focal curves are shown in magenta. Black lines outline boundary cones between TFCDs and spheres. (b) One TFCD bounded by a right circular cone $C$ with apex $P$ and base $E$. (c) The TFCD now occupies the space between two cones $C$ and $C'$, both containing the circular focal curve $E$ but with different apices $P$, $P'$. Two different families of concentric spherical regions are centered at $P$, $P'$, respectively.  (d) A smaller circle $E'$ defines the base of a third cone $C''$ as well as the focal curve of a second TFCD bounded by this  cone. (e)  This construction works also with Type-II FCDs. Black lines outlining boundary cones are continued past the cone apex to the focal circles as gray lines.  } 
	\label{DB_FCDs}
\end{figure}

In addition to the substrate and air interfaces, we might expect that anchoring on the colloid would also affect FCD formation. However, when the colloids were treated with polyvinyl alcohol (PVA) to replace homeotropic anchoring with strong degenerate planar anchoring on their surfaces, qualitatively similar flower textures were observed. We therefore conclude that the importance of the colloid in producing the flower texture lies in the colloid's wetting chemistry that deforms the LC-air interface,  {\em not} the liquid crystalline anchoring on the colloid surface.

In System B,  flowers textures also form in small planar-aligned islands covered by perfluorosilane treated $\mathrm{SiO_2}$ nanoparticles  surrounded by hybrid-aligned, nanoparticle-depleted regions, as shown in Figure \ref{AH_FCDs}b. Here, the elliptical focal curves are visible at the top interface; the hyperbol\ae~extend downward. Figure \ref{AH_FCDs}c shows a less well-ordered flower texture, with striations in the nanoparticle arrangement revealing the arrangement of smectic layers. The ellipse focus of each FCD, where the hyperbola meets the LC-nanoparticle interface, is clearly visible as the center of a set of concentric circles in this plane. The hyperbol\ae~are consistently oriented outward from the center of the planar-aligned region, toward the boundary with the hybrid aligned region. 

The cause of the flower texture in this system is made clear by a cross-sectional SEM image of the smectic layers (Fig.~\ref{AH_FCDs}d). Layers in the planar-aligned region bend outward toward the hybrid-aligned region. Consequently, the layer geometry as viewed from above consists of a planar-anchoring horizontal surface at the top and what can be thought of as a tilted homeotropic surface below, created by the bent layers. This is the geometry that produced a flower texture in Fig.~\ref{MG_flower1}, only upside-down! To see this connection more clearly, we illustrate the geometry of System B schematically in Fig.~\ref{AH_FCDs}e. There, an arbitrarily selected bent layer, colored green, conceptually plays the same role as the curved homeotropic interface of System A. A slight difference between the two experiments is that as distance from the center of the flower texture increases, the typical domain size increases in System B whereas it decreases in System A.

Can these flower textures be described by the law of corresponding cones? In Ref. \cite{meyer2009focal}, the authors suggest that the {\em virtual} branches of the hyperbol\ae~meet at a common point above the center of the flower, rather than the physical hyperbol\ae~as in fan textures under the LCC. We put forth a geometric construction that applies the LCC\ to hybrid-aligned smectics where the homeotropic interface is a surface of revolution, while the degenerate planar interface is flat. We then compare the predictions of this model to the data. 

We begin by reviewing the law of corresponding cones for a family of FCDs. The $i$th FCD is bounded by a cone $C_i$ that has its apex $P$ on the hyperbola $H_i$ and that includes the ellipse $E_i$. Thus, $C_i$ consists entirely of generators, which are normal to the smectic layers and which each connect a point on $E_i$ to a point on $H_i$. In the fan texture, when two ellipses $E_1$ and $E_2$ are tangent at a point $Q$, then their boundary cones $C_1$ and $C_2$ are tangent along an entire generator if $C_1$ and $C_2$ have a common apex $P$ where the hyperbol\ae~$H_1$ and $H_2$ intersect (Fig.~\ref{DB_FCDs}a). Tangency along the generator $\overline{QP}$ means that the layer normals of the two FCDs agree precisely where the FCDs come into contact with each other. Similarly, an FCD with bounding cone apex at $P$ may be joined smoothly onto a family of concentric spheres centered at $P$  (Fig.~\ref{DB_FCDs}b) \cite{Sethna1982}. This is fortunate because concentric spheres, like FCDs, have a focal set of dimension less than two, avoiding energetically costly cusp wall defects.

In the case of tilt grain boundaries split into FCDs, the cone apex $P$ is moved off to infinity along $H$ so that in place of a ``background'' texture of concetric spheres, the FCD matches smoothly onto a background texture of planes whose normal direction matches the hyperbola's asymptotic direction \cite{kleman2000grain}. The bounding cone $C$ has become a bounding cylinder. 

Exactly the same reasoning would apply in a hybrid-aligned smectic if the homeotropic interface were a tilted plane, with the  degenerate planar interface replacing the tilt grain boundary. Thus, if the homeotropic interface is gently curved, it is reasonable to expect the eccentricity to increase with the slope of the interface, which is indeed the case as seen in Fig.~\ref{MG_flower1}. Quantitatively, however, the tilt grain boundary model disagrees with the FCD eccentricities $e$ in our System A: For a (local) homeotropic interface slope of $dh/dr$, the formula given in Ref. \cite{kleman2000grain} predicts
\begin{equation} e^2 = \frac{(dh/dr)^2}{1+(dh/dr)^2} \label{eq:KLe} \end{equation}
which does not fit our data (Fig.~\ref{MG_flower1}e). The curvature of the interface is therefore an important factor.
 But with a curved interface, what is the background geometry into which we are to imagine placing FCDs?
 
We construct such a background texture as follows. First consider a single toric FCD (TFCD), bounded by a right circular cone $C$ with apex $P$, whose base $E$ (the circular focal curve of the TFCD) has radius $a$. As already noted, the layers of the TFCD join smoothly onto a family of spheres concentric about $P$ that exist outside of $C$ (Fig \ref{DB_FCDs}b). But we can also bound a TFCD inside the space between two cones $C$ and $C'$, with the same base but with different apices $P$ and $P'$, respectively. Then, the TFCD also matches smoothly onto spherical layers concentric about $P'$ that exist only inside of $C'$  (Fig \ref{DB_FCDs}c). But once we have this second family of concentric spheres, it is straightforward to cut out from these spheres a cone $C''$ sharing the apex $P'$ with $C'$ but with circular base $E'$ of radius $a'<a$,  and then fill in this cone with a second TFCD  (Fig \ref{DB_FCDs}d). We could continue in this fashion, dividing a region into arbitrarily many nested alternating concentric sphere families and TFCDs, separated by conical bounding surfaces that alternately share a common base or a common apex with the next cone. By construction, axial symmetry is preserved. 

\begin{figure}
	\centering
	\includegraphics[width=.5\textwidth]{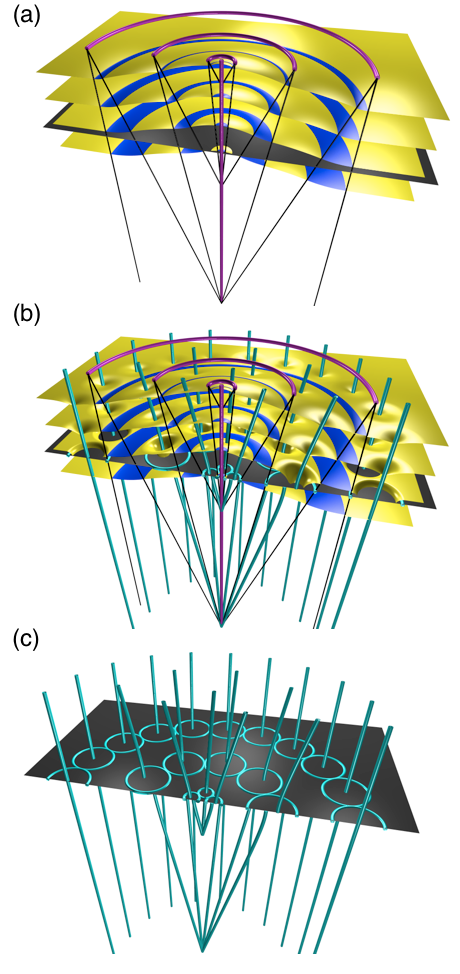}	
	\caption{ (Color)  Flower texture in the LCC. (a) Background texture approximating a curved homeotropic interface and its parallel layers by alternating TFCDs (blue) and concentric spheres or planes (yellow). TFCD focal curves are in magenta. The gray plane, arbitrarily placed, schematically represents the degenerate planar interface, below which there is no smectic physically. (b) FCDs of radially varying eccentricity punctuate the concentric spherical and planar regions. The focal curves of these FCDs are shown in cyan. Focal hyperbol\ae~are physical above the substrate and virtual below.  (c) The focal curves of the smectic flower geometry in (b).}
	\label{DB_FCDs2}
\end{figure}

Now, by turning the picture upside down, we see that an arbitrary surface of revolution can be approximated by an outermost layer of this alternating set of TFCDs and concentric spheres  (Fig.~\ref{DB_FCDs2}a). Thus, if the homeotropic interface is a surface of revolution, this construction generates a smectic layer that approximates that surface, as well as a family of parallel, equally spaced layers below. If the homeotropic interface is planar at large radius, then the background TFCD of largest radius can be made to match smoothly onto planar layers by moving the apex of the largest bounding cone off to infinity, turning the cone into a cylinder. This is the case in Fig.~\ref{DB_FCDs2}. Finally, in each family of concentric spheres bounded between two cones with common apex $P_i$, we can place a ring of FCDs with nonzero eccentricity, all bounded by smaller cones with common apex at $P_i$  (Fig.~\ref{DB_FCDs2}b). These FCDs obey the LCC and match smoothly onto the background texture. Furthermore, because the FCDs' bounding cones have apex below the degenerate planar interface rather than above the homeotropic interface, all of the hyperbolic focal curves will be oriented radially outward (Fig \ref{DB_FCDs2}c)! The virtual branches of the hyperbol\ae, meanwhile, intersect at the common apex $P_i$ below the center of the flower for all FCDs in the same ring. Thus the defining feature of the smectic flower texture can be brought into accordance with the LCC using this construction. 

Note that there is no requirement that the circular focal curves sit at the same height, though we have made this choice in Figs. \ref{DB_FCDs} and \ref{DB_FCDs2} for simplicity. A reasonable, though not unique, alternative is to position each circle at the local center of curvature of the interface's profile. Also, an analogous geometric construction employs Type-II FCDs (Fig \ref{DB_FCDs}e), which produce an interface of purely positive Gaussian curvature   \cite{Bouligand1972, boltenhagen1992focal}, {\em i.e.}, the tilt angle increases with increasing radius. In this case, the picture does not need to be inverted to produce a flower texture, as the cones bounding the concentric spherical regions open upward in Fig \ref{DB_FCDs}e. Finally, we note that this model naturally extends a previous model for polygonal domains, in which the FCDs visible in the experiment are grouped inside background textures of concentric spheres, the various sphere families being glued together by portions of invisible FCDs whose focal curves lie outside the sample \cite{Lavrentovich1986, bragg1934liquid}. 

Does this geometric construction describe the experimental results? While we have successfully captured the radial divergence of hyperbolic focal curves within the LCC, we pause to note some other implications of the model. First, while FCDs within each ring are tangent to their neighbors, the FCDs in different rings must have some space between their ellipses, where the background TFCD interposes between concentric sphere families. This gap is also present in the previously mentioned model for polygonal domains \cite{Lavrentovich1986, bragg1934liquid}. While such a gap is not visible in the experimental images, it can be made small in the model by appropriate choices of bounding cones, to squeeze the background TFCDs into very small angles.

Second, a corollary of the previous point is that FCDs are predicted to pack within each ring but not to show any consistent organization from one ring to the next. This is plausibly consistent with the results in System A (Fig.~\ref{MG_flower1}), provided that the radius of the ring is given some leeway to vary so as to allow for compromise with the quasi-hexagonal packing of ellipses preferred by the degenerate planar substrate. However, System B appears to show FCDs grouped into radial wedges rather than concentric rings (Fig.~\ref{AH_FCDs}), contrary to our expectation from the LCC. Instead, it is possible that each wedge of FCDs in System B has a background texture with no curvature in the azimuthal direction, this curvature being concentrated instead into small-angle tilt grain boundaries between neighboring wedges. 

Thin-film smectics need not precisely obey the LCC because more general layer structures, other than spheres and FCDs, don't incur prohibitive energy penalties if they are generated by virtual focal sets lying outside the sample, {\em i.e.},~their geometry does not require cusps in the smectic \cite{Honglawan2013}. The model presented here demonstrates that the LCC is flexible enough to account for the basic features of the smectic flower but probably not capable of quantitatively describing the textures we observe.

We have demonstrated that control over the orientation mismatch between the hybrid aligning interfaces of a  smectic thin film provides control over the FCD eccentricity and, thus, over complex patterns of self-organization in a liquid crystal. In particular, we have produced smectic flower textures in two experiments: one in which the LC-air interface is curved by pinning at the surface of a colloidal inclusion, the other in which the boundary between a planar-aligned region covered by $\mathrm{SiO_2}$ nanoparticles and a hybrid-aligned region exposed to air creates effective tilted hybrid anchoring. The radial outward orientation of the focal hyperbol\ae, unique to the flower texture, is deduced to arise from the outward tilt of the homeotropic interface's normal vector. We have extended previous LCC-based models by proposing a geometric model for hybrid-aligned smectics in the case that the homeotropic interface is a surface of revolution. The resulting nested system of background TFCDs and concentric spheres naturally allows for FCDs with their hyperbol\ae~oriented radially outward, as in the  flower texture. Comparing the implications of this model with the experimental results shows that the LCC can accomodate the arrangement of FCDs in a flower texture but does not describe the details of their packing behavior. The findings presented here will open the door to crafting highly sophisticated self-assembled patterns in smectic liquid crystals, for use in guiding functional materials such as colloids and nanoparticles into technologically useful arrangements, by clever preparation of the boundaries.

\acknowledgements
This work was supported in part through the UPenn MRSEC Grant NSF DMR11-20901. DAB was supported by NSF Graduate Research Fellowship DGE-1321851. DAB and RDK thank the Isaac Newton Institute for its hospitality while this work was completed.  This work was partially supported by a Simons Investigator award from the
Simons Foundation to Randall D.~Kamien.

\appendix*
\section{Experimental Details}
System A is obtained by dispersing 0.1\% wt. of silica beads of nominal diameter $5\; \mu\mathrm{m}$ (Polysciences, Inc.) in 4-n-octyl-4'-cyanobiphenyl (8CB purchased from Kingston Chemicals Limited), which displays a SmA phase between 22.3$^\circ$C and 33.4$^\circ$C. The solid particles are treated to induce strong homeotropic anchoring on 8CB using the same chemistry described in reference \cite{Gharbi2011anchoring}. Hybrid-aligned thin films (degenerate planar anchoring at the base and homeotropic anchoring in contact with air), of thickness smaller than the dimension of particles, are prepared by drawing a small amount of SmA/particles suspension across holes of diameter 600 $\mu$m fabricated using standard lithographic techniques from KMPR negative photoresist (Microchem Corp.) on copper substrates. \\
\\
The profile of the SmA interface in contact with air was characterized using intereferometric technique: scanning white-light interferometry (SWLI). The measurements were taken by a Zygo NewView 6200 interferometer. The experimental system was studied under an upright optical microscope (Zeiss AxioImager M1m) in transmission mode equipped with a heating stage (Bioscience Tools, temperature regulated at 0.1 $^\circ$C) and a set of crossed polarizers. Images were recorded with a high-resolution camera (Zeiss AxioCam HRc) and high-speed camera (Zeiss AxioCam HSm). \\
\\
System B is produced via coassembly of fluorosilanized silica nanoparticles (NPs) and the semi-fluorinated smectic liquid crystal (LC), (4'--(5,5,6,6,7,7,8,8,9,9,10,10,11,-11,12,12,12-heptadecaflu-orododecyloxy)-biphenyl-4-carboxylic acid ethyl ester) \cite{Honglawan2011}. The silica NPs ($d = 100 \pm 3$ nm, 30 wt \% in isopropanol from Nissan Chemicals) were functionalized with (heptadecafluoro-1,1,2,2,-tetrahydrodecyl) dimethylchlorosilane (HDFTHD) (99\%) (Gelest, Inc.) by following the literature \cite{xu2012transparent}.  The experiment required the mixture of LC and the silanized NPs at weight ratio of 9:1 dispersed in the fluorinated solvent, Novec 7300, provided by 3M at 1 wt \%. The mixture, drop-casted onto a clean Si-wafer was heated on a Mettler FP82 and FP90 thermo-system hot stage to the isotropic phase at 200$^\circ$C,  cooled down to smectic temperature at 114$^\circ$C at 5$^\circ$C/min to form a smectic flower texture confined by two planar anchoring interfaces covered by NPs, and subsequently quenched to room temperature. The NPs are dispersed in the isotropic phase but, upon cooling to the smectic phase, they separate from the LC to densely cover the two interfaces, as illustrated schematically in Fig \ref{AH_FCDs}a. \\
\\
The quenched structures from the coassembly were characterized by scanning electron microscopy (SEM) on FEI Strata DB235 focused ion beam (FIB) system at 5 kV. Molecular anchoring of LC on NPs was determined by Zeiss Axiolmager M1m upright microscope with crossed polarizers (on a Minus K Vibration Isolation Platform with monochrome camera: AxioCam HSm, and color camera: AxioCam HRc) through reflectance mode. 

\bibliographystyle{h-physrev}


\end{document}